\newif\ifarxiv
\arxivtrue   

\ifarxiv
  \makeatletter
  \def\input@path{{acm-templates/}}
  \makeatother
  \documentclass[manuscript,screen]{acmart} 
  \renewcommand\footnotetextcopyrightpermission[1]{}

\else
    \documentclass[acmsmall,screen,review,anonymous]{acmart}
\fi

\usepackage{amsmath}
\usepackage{algorithmic}
\usepackage{graphicx}
\usepackage{textcomp}
\usepackage{xcolor,colortbl}
\usepackage{courier}
\usepackage{booktabs}
\usepackage[T1]{fontenc}

\usepackage{tcolorbox}
\usepackage{multirow}
\usepackage{array}
\usepackage{enumitem}
\usepackage{etoolbox}

\usepackage[utf8]{inputenc}

\usepackage{ulem}
\usepackage{url}
\usepackage{hyperref}
\hypersetup{
    colorlinks,
    linkcolor={black!100!black},
    citecolor={black!100!black},
    urlcolor={blue!40!black}
}

    \def\BibTeX{{\rm B\kern-.05em{\sc i\kern-.025em b}\kern-.08em
    T\kern-.1667em\lower.7ex\hbox{E}\kern-.125emX}}

\AtBeginDocument{%
  \providecommand\BibTeX{{%
    Bib\TeX}}}

\ifarxiv
    \renewcommand\footnotetextcopyrightpermission[1]{}
\else
    \setcopyright{acmlicensed}
    \copyrightyear{2025}
    \acmYear{2025}
    \acmDOI{XXXXXXX.XXXXXXX}
    \acmConference[Conference acronym 'XX]{Make sure to enter the correct
      conference title from your rights confirmation email}{June 03--05,
      2025}{Woodstock, NY}
    \acmISBN{978-1-4503-XXXX-X/2018/06}
\fi

\begin{document}

\definecolor{paleyellow}{rgb}{1, 1, 0.85}
\definecolor{lower}{rgb}{0.88235,0.7451,0.41569}
\definecolor{lower}{rgb}{0.94, 0.85, 0.6}
\definecolor{higher}{rgb}{0.5, 0.8, 0.75}
\definecolor{rowColor}{rgb}{0.937, 0.937, 0.937}
\definecolor{mauve}{rgb}{0.25,0,0.52}
\definecolor{darkerteal}{RGB}{1, 77, 78} 
\newcommand{\FIXME}[1]{\textcolor{red}{Revision: \uline{#1}}}
\newcommand{\fixme}[1]{#1}
\newcommand{\MLE}[1]{\textcolor{blue}{MLE: #1}}
\newcommand{\mle}[1]{\textcolor{blue}{MLE: #1}}
\newcommand{\Sarah}[1]{\textcolor{red}{Sarah: #1}}
\newcommand{\sarah}[1]{\textcolor{red}{Sarah: #1}}
\newcommand{\veronica}[1]{\textcolor{green}{Veronica: #1}}
\newcommand{\Veronica}[1]{\textcolor{green}{Veronica: #1}}
\newcommand{\peggy}[1]{\textcolor{orange}{Peggy: #1}}
\newcommand{\Peggy}[1]{\textcolor{orange}{Peggy: #1}}
\newcommand{\Chris}[1]{\textcolor{brown}{Chris: #1}}
\newcommand{\chris}[1]{\textcolor{brown}{Chris: #1}}

\newcommand{\subheading}[1]{\vspace{2pt}\noindent\textbf{#1}}
\newcommand{\qualquote}[2]{%
  \textit{\textcolor{mauve}{``#1}''} \hspace{0em}%
  \textnormal{(#2)}%
}

\DeclareRobustCommand{\code}[1]{%
  \textit{\textcolor{darkerteal}{#1}}%
}

\newcommand{\GroupOne}[0]{\section{Internal Problem-Solving}}
\newcommand{\GroupTwo}[0]{\section{Hitting Targets}}
\newcommand{\GroupThree}[0]{\section{Environmental Disruptions}}
\newcommand{\GroupFour}[0]{\section{Job Fit and Social Aspects}}

\newcommand{\summarytcolorbox}[1]{
\begin{tcolorbox}[colback=paleyellow, colframe=black, colbacktitle=white, coltitle=black, boxsep=0pt, top=3pt, bottom=3pt, left=6pt, right=6pt, before skip=3pt]
    \noindent #1
\end{tcolorbox}
}

\newcolumntype{L}[1]{>{\raggedright\arraybackslash}m{#1}}

\newcolumntype{R}[1]{>{\raggedleft\arraybackslash}m{#1}}

\newcolumntype{C}[1]{>{\centering\arraybackslash}m{#1}}

\makeatletter
\patchcmd{\@makecaption}
  {\scshape}
  {}
  {}
  {}
\long\def\@makecaption#1#2{%
  \vskip\abovecaptionskip
  \small 
  \sbox\@tempboxa{#1. #2}%
  \ifdim \wd\@tempboxa >\hsize
    #1. #2\par
  \else
    \global \@minipagefalse
    \hb@xt@\hsize{\hfil\box\@tempboxa\hfil}%
  \fi
  \vskip\belowcaptionskip}
\makeatother

\newenvironment{quoteblock}%
  {\begin{list}{}{\leftmargin=2em \rightmargin=1em \topsep=5pt \parsep=0pt \itemsep=8pt} \item[]}%
  {\end{list}}


\title{Good Vibrations? A Qualitative Study of Co-Creation, Communication, Flow, and Trust in Vibe Coding}
\author{Veronica Pimenova}
\affiliation{%
  \institution{University of Michigan}
  \city{Ann Arbor}
  \state{Michigan}
  \country{USA}
}
\email{pimenova@umich.edu}

\author{Sarah Fakhoury}
\affiliation{%
  \institution{Microsoft Research}
  \city{Redmond}
  \state{Washington}
  \country{USA}
}
\email{sfakhoury@microsoft.com}

\author{Christian Bird}
\affiliation{%
  \institution{Microsoft Research}
  \city{Redmond}
  \state{Washington}
  \country{USA}
}
\email{cbird@microsoft.com}

\author{Margaret-Anne Storey}
\affiliation{%
  \institution{University of Victoria}
  \city{Victoria}
  \state{British Columbia}
  \country{Canada}
}
\email{mstorey@uvic.ca}

\author{Madeline Endres}
\affiliation{%
  \institution{University of Massachusetts Amherst}
  \city{Amherst}
  \state{Massachusetts}
  \country{USA}
}
\email{mendres@umass.edu}

\renewcommand{\shortauthors}{Pimenova et al.}

\begin{abstract}

Vibe coding, a term coined by Andrej Karpathy in February 2025, is a compelling, controversial natural language programming paradigm in AI-assisted software development. Centered on iterative co-design with AI, vibe coding emphasizes flow and experimentation over specification. While initial studies explore this paradigm, most focus on code artifacts or theories with limited empirical backing. There remains a need for a grounded understanding of vibe coding as perceived and experienced by developers. We analyze 134 posts and over 5,000 practitioner comments on LinkedIn and Reddit to understand their experiences, complemented by 11 developer interviews. We characterize what vibe coding is, why and how developers use it, where it breaks down, and which emerging practices support it. Grounded in qualitative analysis, we present a theory of vibe coding centered on conversational interaction with AI, co-creation, and developer flow and joy. Our theory captures how trust in AI is regulated from delegation to co-creation, and how trust supports the developer experience by sustaining flow. We surface recurring pain points in areas including specification, reliability, debugging, latency, code review burden, and collaboration, and present best practices reported to mitigate these challenges. We conclude with implications for future development of AI tools and research directions.

\end{abstract}


\received{20 February 2007}
\received[revised]{12 March 2009}
\received[accepted]{5 June 2009}

\maketitle

\section{Introduction}


\begin{quoteblock}
    \qualquote{It’s not about chaos. It’s about flow writing code in a rhythm where your mind is free to create, unburdened by boilerplate.}{L46}
\end{quoteblock}

The phrase ``vibe coding''  was first introduced by AI researcher Andrej Karpathy in February of 2025 and immediately became popular because it sparked imagination and captured the enthusiasm of a new style of accelerated creative programming using LLMs that frees the developer from the usual constraints of thinking about technical and engineering details before getting into the ``vibes''.  The term, however, has been criticized~\footnote{See Hacker News for a discussion around this criticism, sparked by comments by Andrew Ng} 
as it may trivialize the importance of AI supported development. 
Despite these criticisms, and a lack of a formal definition or agreement of what this term means, the phrase is frequently used and may be in part responsible for steering a shift toward a new paradigm of programming that is less reliant on formalizing requirements before coding.  

The timing of this framing has also coincided with an industry movement towards improving the ``experience'' of developers. This movement recognizes that joy while programming can lead to software solutions that are not only more creative~\cite{graziotin, greiler2022actionable}, but also bring delight to the customers of that software. The vibe programming paradigm also coincides with extensive research to understand the impact of Gen AI on development velocity and software quality. 

The speed of this unique paradigm shift warrants an investigation to understand how vibe coding is evolving and what this may signal for the future of software development and engineering practices. Our timely research seeks to understand what a community of developers perceive to be true when they use this phrase, why experts and non developers vibe code, and how they do so.  We also seek to uncover perceptions about the software they create, and the challenges they face while doing so. 
In addition, we investigate the perceived risks of abandoning a traditional, principled, design-forward approach to software engineering. Specifically, we investigate the implications of when developers embrace LLMs to rapidly develop solutions they may not have anticipated when they sat down to vibe code. These early signals from developers can help practitioners and researchers recognize these emerging risks and plan solutions for measuring or mitigating these risks, and bring more awareness to these risks across our field.


To arrive at a theory that describes developer perceptions towards ``vibe coding'', we code social media posts made by practitioners adopting vibe coding techniques, and augment these early insights with in-depth semi-structured interviews. Our rational for analyzing social media posts is that the channels we consider capture communication from a ``community of practice''~\cite{wenger1998communities} that has formed around vibe coding on Reddit and LinkedIn. 
On Reddit, the community has grown to 159K members, and posters describe their experiences, share their concerns and best practices, and seek feedback on the products  they ``vibe code''. For our analysis, we scraped Reddit posts from \texttt{r/vibecoding}
and LinkedIn, where we searched for recent and popular posts tagged with \texttt{\#vibecoding}. We further conducted 11 interviews with practitioners to gain deeper practitioner insights, including challenges and experiences. In total, we use a flexible qualitative methodology~\cite{deterding2021flexible} to analyze over 5,000 comments (14,939 sentences) made across 134 posts on vibe coding experiences and practice. 

Our integrated analysis leads to a \textbf{theory} that connects how vibe coding is seen by many as a new \textbf{interaction paradigm for software development},  that supports the \textbf{co-creation of software by humans and AI agents}, and captures the \textbf{psychological flow experiences of developers while ``vibing''}.  We further capture the \textbf{pain points}, developers encounter, and the \textbf{practices} they have adopted to address these pain points. We also show that \textbf{trust} in AI mediates the co-creating experience and impacts vibe coding flow. Finally we share the main \textbf{risks} that developers are concerned about in the vibe coding emerging community of practice. Succinctly, our paper leads to the following main four contributions:
\begin{enumerate}
  \item A \textbf{definition} and empirically grounded \textbf{theory} of vibe coding that connects interaction, co-creation, flow, and trust.
  \item A \textbf{characterization of practice}: where and why vibe coding happens, what developers delegate, and how they guide models.
  \item A synthesis of \textbf{pain points and strategies}, including communication, planning/abstraction, verification, code quality, and collaboration concerns, and strategies to mitigate these concerns.
  \item \textbf{Implications} for tools and research: guidelines and affordances to support flow while taking reliability and team coordination into account.
\end{enumerate}
 

\section{Background}

\label{sec:Background}

To provide context for our study, we overview background relating to natural language programming paradigms (Section~\ref{subsec:BackgroundNLCoding}), AI co-creation (Section~\ref{subsec:BackgroundOtherAICoding}), and flow (Section~\ref{subsec:BackgroundFlow}).

\subsection{Natural Language Programming Paradigms}
\label{subsec:BackgroundNLCoding}

We use \emph{natural language programming} to denote approaches where developers use everyday language to \textit{specify}, \textit{modify}, or \textit{orchestrate} software, drawing on HCI work on “natural programming” that examines how people express computational intent before learning formal syntax~\cite{myers2004natural}.

The ambition to program with natural language dates to early computing. 
Business-oriented languages such as COBOL adopted English-like syntax to broaden access~\cite{cobol1960}. 
The fourth-generation language (4GL) movement pushed toward declarative, task-level specification 
promising ``programming without programmers'' via high-level abstractions for data and business logic~\cite{harel2002biting}. 
Pure natural language, however, proved insufficient without \emph{grounding}, connecting language to executable semantics within well-defined domains. 
Classic AI systems grounded language in narrow, executable domains such as SHRDLU in a blocks world that turned commands into actions and explanations~\cite{winograd1972}, and LUNAR by translating Apollo geology questions into structured database queries with checkable results~\cite{woods1973}.

Two complementary strategies emerged to manage ambiguity. 
\emph{Controlled natural languages} (e.g., Attempto Controlled English) restrict grammar and vocabulary so sentences map deterministically to logic~\cite{fuchs1996ace,fuchs2008ace}. 
In parallel, \emph{programming by demonstration} (PBD) and \emph{by example} (PBE) infer intent from concrete artifacts~\cite{cypher1993watch}, later formalized by program synthesis methods like \textsc{FlashFill}~\cite{gulwani2011flashfill} that solve for programs from I/O pairs or partial specifications~\cite{solar2008sketch}.
Collectively, these show that natural language works best when paired with examples, types, tests, or partial programs.

Machine learning shifted the field from rule-based translation to learned, data-driven mappings. 
Early statistical semantic parsers mapped utterances to logical forms with probabilistic models~\cite{zettlemoyer2005,zettlemoyer2007}. 
Deep learning then enabled sequence-to-sequence NL$\rightarrow$code trained on parallel corpora, performing best where \emph{executable oracles} provide objective verification.  
For example, text-to-SQL on Spider~\cite{yu2018spider}, Python snippets in CoNaLa~\cite{yin2018conala}, and model evaluations like Codex on HumanEval~\cite{chen2021codex}. 
Pairing neural flexibility with deterministic checks proved crucial.

Large language models trained on code broadened scope and interaction. 
Systems like Codex moved beyond domain-specific mappings to general-purpose languages~\cite{chen2021codex}, enabling \emph{conversational programming} where developers iteratively describe goals, request modifications, and refine implementations in natural language~\cite{ross2023programmer}. 
This elevates language from a translation interface to a primary programming surface, but without systematic specification or testing, iterations can drift and compound errors.

We organize natural language programming along three design axes:
\begin{itemize}[leftmargin=1.25em, itemsep=2pt, topsep=2pt]
  \item \textbf{Interaction mode:} one-shot translation vs.\ conversational refinement.
  \item \textbf{Grounding strategy:} pure NL vs.\ NL augmented with artifacts (I/O examples, tests, types, schemas, partial programs).
  \item \textbf{Verification:} ad-hoc inspection vs.\ systematic oracles (unit/property tests, expected outputs, formal specs).
\end{itemize}
These axes clarify trade-offs between fluid interaction and precise specification, naturalness and semantic grounding, and creative exploration and correctness guarantees.

Different paradigms occupy distinct regions of this space (Table~\ref{tab:nl-paradigms-revised}). 
NL$\rightarrow$code systems often use one-shot interaction with limited grounding and rely on manual verification~\cite{chen2021codex}. 
NL$\rightarrow$DSL~\cite{yu2018spider,lin2018nl2bash} and PBE~\cite{gulwani2011flashfill} trade expressiveness for reliability via schema/DSL constraints or executable oracles. 
Conversational environments emphasize high-interaction dialogue~\cite{ross2023programmer}, progressively building grounding via context~\cite{barke2023grounded}, though verification often remains informal~\cite{vaithilingam2022expectation}.

\begin{table}[t]
\centering
\footnotesize
\begin{tabular}{@{}lccc@{}}
\toprule
\textbf{Paradigm} & \textbf{Interaction} & \textbf{Grounding} & \textbf{Verification} \\
\midrule
NL$\rightarrow$Code (Codex, Copilot) & One-shot/Limited & Low–Medium & Manual/Ad-hoc \\
Conversational (ChatGPT, Claude) & High/Dialogue & Medium (context) & Progressive/Informal \\
NL$\rightarrow$DSL (Spider, NL2Bash) & One-shot & High (schema/DSL) & Executable oracle \\
PBE (FlashFill, Sketch) & One-shot (via examples) & High (I/O pairs) & Oracle (I/O pairs) \\
\textbf{Vibecoding (This Work)} & \textbf{High/Conversational} & \textbf{Low/Minimal} & \textbf{Ad-hoc/Improvisational} \\
\bottomrule
\end{tabular}
\caption{Natural language programming paradigms characterized by our three axes. Vibecoding prioritizes interaction fluidity over formal grounding and verification.}
\label{tab:nl-paradigms-revised}
\end{table}

Vibe coding occupies the high-interaction, low-grounding corner: developers iterate in dialogue with minimal formal scaffolding, relying on improvisational verification and run–refine loops.
This aligns with exploratory programming where goals evolve through programming~\cite{kery2017exploring}. 
Sections~\ref{sec:conversation}–\ref{sec:trust} examine how practitioners navigate these trade-offs, what this affords for rapid progress, and where it strains reliability.

\subsection{Other Paradigms of AI Co-Creation}
\label{subsec:BackgroundOtherAICoding}

Work on AI-assisted programming and design often makes user intent explicit and interpretable, in contrast to the intuition-driven style of vibecoding. 
Building on instrumental interaction theory~\cite{beaudouin2000instrumental}, Riche \emph{et al.} advocate reified controls—sliders, prompt fragments, and other manipulables—that let users externalize, adjust, and reuse intentions rather than iteratively retyping prompts~\cite{riche2025ai}.

Studies also underscore the value of structured collaboration. 
Designers report AI is most useful as an ideation partner that expands the solution space~\cite{khan2025beyond}. 
In programming, proactive assistants such as \textsc{Codelleborator} improve efficiency when help is timely and context-aware, but poor timing harms flow~\cite{pu2025assistance}. 
Co-creation quality thus depends not only on what the AI generates but on when and how it is invoked.

A complementary line of research treats code as a medium for exploration. 
\textsc{Pail} elicits and tracks goals while surfacing implicit LLM decisions, using LLM-synthesized, executable prototypes as disposable sketches to test alternatives and trade-offs~\cite{zamfirescu2025beyond}. 
\emph{Code Shaping} enables direct manipulation of code via free-form sketches, arrows, pseudocode, and natural language on or around the source~\cite{yen2025code}. 
Together, these systems frame co-creation as iterative design that moves between problem and solution spaces, privileging executable sketches and direct manipulation over one-shot generation.

Other paradigms suit non-developers such as end-users and scientists. 
In a “specify-and-verify” workflow, users ground intent with concrete input/output examples and steer solutions by running code against those examples~\cite{pickering2025humans}.
Strong examples improved task intent and code quality more than programming expertise, and outperformed model-driven clarification.
Field observations of scientists show similar practices, using LLMs like searchable documentation to explore APIs, then verifying via run-and-inspect with emphasis on correctness and reproducibility~\cite{obrien2025scientists}. 
Both lines illustrate co-creation through concretized intent and empirical checking.

These approaches emphasize reproducibility and clear specification. 
Vibe coding takes a different tack, favoring rapid, minimally formal iteration where users adjust only when outcomes diverge from goals. This immersive, flow-oriented style is the focus of our investigation.

\subsection{Flow and Software Development}
\label{subsec:BackgroundFlow}

Early reports by vibe coders describe feelings of a flow state while ``vibing'' and although we didn't start our study with the intention to study flow, flow and feelings of joy emerged as salient aspects of how developers feel and what motivates them to vibe code.  The original theory of flow lays out a rich set of \emph{characteristics} and key \emph{conditions} for flow.  We use these constructs to interpret how vibe coders interact and co-create with AI in our analysis (described later in our paper).
%
Csikszentmihalyi coined the term flow to describe the optimal psychological experience humans feel when they are immersed in an activity that is \emph{intrinsically rewarding}~\cite{csikszentmihalyi1990flow}. Flow occurs when people \emph{lack self-consciousness} of their actions, \emph{lose track of time}, but feel they are more in \emph{control} of tasks that feel \emph{effortless} with minimal conscious effort needed to guide decisions~\cite{csikszentmihalyi1990flow}.

For flow to occur, certain \emph{conditions} are important~\cite{csikszentmihalyi1990flow}. Tasks need to be \emph{challenging} enough to increase a person's skills but not cause stress, and for a person to feel in control they need to feel confident in their capabilities and receive ongoing \emph{feedback} on progress towards clear \emph{intermediate goals}. 
%
Flow experiences are enhanced when shared with others who collectively believe and trust in their shared capabilities to be creative~\cite{csikszentmihalyi1990flow, Salanova2014}.  Collective flow experiences enhance creativity, learning, and motivation~\cite{csikszentmihalyi1996creativity}. 

 In software development, developers enter a flow state when immersed in coding or debugging activities, and is an important aspect of the developer's experience~\cite{greiler2022actionable}.   %
SE researchers have studied a number of barriers to flow such as interruptions~\cite{meyer2017,ritonummi2024exploring,ma2024breaking}, distractions~\cite{meyer2019fostering}, lack of focus time~\cite{ritonummi2024exploring,brown2023developer,khemka2024toward}, tool friction~\cite{ritonummi2024exploring,brown2023developer}, or getting stuck while waiting on others~\cite{ritonummi2024exploring}. 
Csikszentmihalyi makes less mention of these or other barriers to flow and instead emphasizes the characteristics and conditions for flow to occur.  
Meyer et al. touch on the importance of goal setting to facilitate flow in development~\cite{meyer2019fostering},   while other researchers touch on how fast feedback may facilitate flow~\cite{petersen2011measuring,noda2023devex_queue}. Finally, Ritonummi et al. describe how not enough (or too much) challenge during development may not lead to flow~\cite{ritonummi2024exploring}.


More recent research on GenAI use in SE has found that developers are more likely to experience flow when using GenAI~\cite{mckinsey2023genai, kalliamvakou2021space, dora2024report, butler2024dear}. 
In our study of vibe coding, we build on this work and find developers eloquently articulate how vibe coding supports the necessary conditions for flow. The natural language interface for vibe coding enhances creativity, and the highly-iterative approach of vibing and reification of early ideas, provides critical feedback to developers of their continuous progress towards intermediate goals. The characteristics and conditions for flow emerge during our analysis of vibe coding interactions and experiences. 





\section{Methodology and Research Questions}

\label{sec:Methods}

Our goal was to develop a theory of vibe coding practice and experience that is \textit{rigorous}, and \textit{generalizable}, while still capturing heterogeneous practitioner perspectives. 
We collected data from public social-media posts (Reddit and LinkedIn) and semi-structured interviews, triangulating across sources to reduce bias and strengthen validity. 
Using a well-established flexible qualitative methodology~\cite{deterding2021flexible}, we used iterative analysis to build consensus between authors. The rest of this section details our data collection, dataset, and analysis methodology. 

\subsection{Guiding Research Questions}
\label{subsec:rqs}

To guide data collection and analysis, we collectively formulated five \textit{initial research questions (RQs)}:

\begin{itemize}[topsep=0pt, partopsep=0pt, parsep=0pt, itemsep=0pt]
    \item \textbf{RQ1--\code{Definition}:} \textit{What} is vibe coding?
    
    \item \textbf{RQ2--\code{Practice}:} \textit{Why} do programmers vibe code and \textit{when} do they do it?
    \item \textbf{RQ3--\code{Perceptions}:} What are the \textit{perceptions} towards vibe coding?
    \item \textbf{RQ4--\code{Pain Points}:} What are the \textit{challenges} and \textit{risks} associated with vibe coding?
    \item \textbf{RQ5--\code{Best Practice}:} What \textit{best practices} are emerging to handle these challenges? 
    
\end{itemize}

\noindent These initial questions were developed via discussions between all authors over a series of weekly meetings from May to August 2025. To aid in question construction, authors engaged in a process of prolonged engagement~\cite{guba1989fourth}, reading hundreds of vibe coding posts across two social media platforms. In qualitative research, research questions serve as orienting tools; they help guide data collection, coding, and interpretation, but they can change over the research process and do not always lend themselves to neat, definitive answers~\cite{agee2009developing}. While we consider all of these questions in our analysis, we do not organize our findings as one-to-one answers. For traceability and coherence, we conclude the results section with a summary in Section~\ref{sec:resultssummary} that maps each research question to the most relevant findings.

\subsection{Data Collection}


\label{subsec:DataCollection}

To both capture generalized perspectives and also permit data triangulation, we collected data from three sources: Reddit, LinkedIn, and semi-structured interviews. These sources represent a mix of public anonymous discussion, public identified posts, and private opinions. Table~\ref{tab:datasources} overviews the size and content of each data source. Throughout this paper, textual quotes are labeled by data source (R for Reddit, L for LinkedIn, or I for Interview) and post number for clarity.

\begin{table}[t]
\footnotesize
\begin{tabular}{lrrr}
\hline
\textbf{Data Source}       & \textbf{Number} & \textbf{Words}  & \textbf{Date Collected} \\ \hline
Vibe Coding Related Reddit Discussions & 46     & 102,741 & May 27, 2025 \\
Vibe Coding Related LinkedIn Posts            & 88     & 25,493 & May 27, 2025 \\
Semi-Structured Interviews            & 11     & 64,650  & June 7--July 10, 2025 \\

\midrule
\textit{Total}                     &   145  & 192,884   
\end{tabular}
\caption{Overview of data sources analyzed in this study. \label{tab:datasources}}
\vspace{-10pt}
\end{table}

\textbf{Reddit Data:} A platform of topic-specific forums with distinct community norms, Reddit is a major venue for programming discussions and has been widely used in academic research on online communities and professional practices~\cite{proferes2021studying,fiesler2024remember}.  In software engineering, Parsons~\emph{et al.} studied Reddit threads to understand developers' privacy concerns, demonstrating how programming communities on Reddit surface perspectives and experiences around real work contexts~\cite{parsons2023understanding}.
The subreddit we studied,\texttt{r/vibecoding}, was created on February 8th, 2025 (just six days after the term was coined), and has already grown to 156,000 members, ranking in the top 1\% of Reddit communities\footnote{\url{https://www.reddit.com/r/vibecoding/}, Reported size as of September, 2025}. Posters share experiences, ask for feedback, and exchange memes about vibe coding. 
Related discussions also appear on forums like \texttt{r/ProgrammerHumor} and \texttt{r/cursor}. 

To capture a range of perspectives, we collected our data set via four Reddit queries: two across all posts on \texttt{r/vibecoding} (filtered by top posts of all time, most recent posts) and two across all Reddit forums via a search for ``vibe coding'' (most relevant posts, top of all time). From each search, we used a new incognito browser and scraped the first 20 results, resulting in 80 threads (posts plus comments). As this process could have resulted in duplicate or irrelevant posts, we manually screened all 80 threads, resulting in a final set of 66. 
We analyzed a random subset until reaching saturation (46 threads, 102,741 words; see Section~\ref{subsec:AnalysisMethodology}). All data was collected on May 27th, 2025. 


\textbf{LinkedIn Data:} 
Unlike Reddit, LinkedIn posts are typically tied to real names and professional identities. This makes LinkedIn a valuable contrasting data source, capturing vibe coding discussions in a reputationally sensitive context. We collected 99 posts by searching for  \texttt{\#vibecoding}: 50 of the posts were the most recent (as of May 27th, 2025) and 49 were top-liked posts. After removing duplicates and manually screening for relevance, we analyzed 88 unique posts (25,493 words).

\textbf{Semi-structured Interviews:} To gain deeper insights on our findings from the social media analysis, we conducted 11 semi-structured interviews with practitioners who had engaged in vibe coding. 
Each interview was structured around our five initial guiding research questions (see Section~\ref{subsec:rqs}), with room left open to follow-up on interesting comments. We performed a pilot interview to finalize our interview script.
Interviews were conducted via Zoom or Microsoft Teams, lasted 30–60 minutes, and were recorded and transcribed. Recruited through snowball sampling and a post on \texttt{r/vibecoding}, participants completed a pre-interview survey on their programming and vibe coding experience. To be eligible, participants had to have vibe coded before. Our sample intentionally included a range of software engineering backgrounds, from novice developers to professionals with over 40 years of experience (Table~\ref{tab:participants}). Interviewees were not compensated, and we stopped after reaching saturation when no new findings were confirmed with respect to our preliminary analysis of the social media data (see~Section~\ref{subsec:AnalysisMethodology}).

\begin{table}[t]
\footnotesize
\centering
\begin{tabular}{@{}llrr@{}}
\toprule
\textbf{Participant ID }& \textbf{Programming Experience} & \textbf{Years of Program-}& \textbf{Vibe Coding Experience} \\ 
 & &  \textbf{ming Experience} & \\ \midrule
I1          & General Software Engineering & 11 & Personal Projects \\
I2          & General Software Engineering & 5 &  Personal Projects \\
I3          & Web Development &  15 &  Data analysis \\
I4          & Data Science & 11 &  Data analysis \\
I5          & Machine Learning  & 15 & Machine learning code at work \\
I6          & Machine Learning & 17 & Developing company products \\
I7          & General Software Engineering & 46 & Personal projects \\
I8          & Back-end Development & 10 & Data analysis \& personal projects \\
I9          & Web Development & 0 & Company website \\
I10         & Machine Learning & 8 & Personal Projects \\
I11         & Machine Learning & 4 & Front end \& data processing \\ \bottomrule
\end{tabular}
\caption{Interviewee programming experience and vibe coding experience. All participants have tried vibe coding, and many regularly vibe code for personal projects or work-related tasks. One participant (I9) had not had programming experience before vibe coding.\label{tab:participants}} 

\end{table}

\subsection{Analysis Methodology}
\label{subsec:AnalysisMethodology}

We followed Detedring and Waters's flexible qualitative analysis approach~\cite{deterding2021flexible}. Adapted from grounded theory, this approach facilitates directed but flexible analysis of a large number of texts ($N$ > 30) using Qualitative Data Analysis (QDA) Software. Though originally developed for interviews, it has been successfully applied to programming-related social media posts~\cite{Newman2025ADHDICSE}, making it well-suited to our study. Our methodology involved three stages: (1) applying top-level index codes related to our research questions, (2) developing and applying analytic low-level codes, and (3) iterative validation and grouping of analytic codes via iterative axial coding. All stages were conducted in the QDA NVivo 15, with Miro boards supporting collaborative axial coding.

\textbf{Analysis Stage 1---Index Codes:} In the first pass, three authors applied seven high-level index codes to broad sections of text. Five were based on our guiding research questions: \code{Definition}, \code{Practice}, \code{Perceptions}, \code{Pain Points}, and \code{Best Practice}. We added an index code for \code{Future Desires} after team discussion. Following Deterding and Waters~\cite{deterding2021flexible}, we also marked particularly concise or evocative passages as \code{Interesting Quote}. While indexing, coders wrote memos that were discussed every week. These memos served as the foundation of the analytic codes applied during stage 2.

\textbf{Stage 2---Developing and Applying Analytic Codes:} Two authors reviewed text within each index code to develop low-level analytic codes, using \textit{in vivo} coding where possible, an approach where participants’ own words become the names of codes~\cite{khandkar2009open}. This approach can result in a large number of low-level codes, so the authors met weekly from May to August 2025 to regroup codes and identify themes. Saturation for a data source was reached when no new codes emerged in a new document.

We did not compute inter-rater reliability (IRR). Instead, we used a negotiated agreement: researchers independently coded subsets of the data, then met to resolve differences and build consensus, repeating this iteratively during codebook development.
As McDonald~\emph{et al.}~\cite{mcdonald2019reliability} argue, IRR is not always appropriate for qualitative research as it can falsely suggest objectivity, discourage interpretive disagreement that strengthens analysis, and is often misaligned with theory-building goals. 
In our study, codes were not treated as fixed ``labels'' for quantification but as analytic tools to surface and refine emerging themes. 
Our emphasis on discussion, memoing, and consensus-building provides methodological rigor while remaining faithful to the interpretive aims of our work.

\textbf{Stage 3---Iterative Axial Coding:} We used \textit{axial coding} to synthesize key themes by merging and connecting analytic codes into abstract categories and conceptual relationships~\cite{vollstedt2019introduction}. We mapped themes to vibe coding practices, end experiences. We used Miro~\cite{miro} for organizing codes and facilitating the axial coding process, enabling effective merging and connecting of codes into abstract categories and conceptual relationships. Reddit data was used to develop the initial codebook and groupings. We then applied this codebook to the interview and LinkedIn data, adding new codes as needed, followed by a final round of axial coding. This approach supported both cross-platform theme triangulation and identification of platform-specific differences.

In this qualitative research, our focus is on investigating diverse perspectives rather than quantifying occurrences. 
As a result, we do not enumerate the frequency of themes across participant responses to avoid implying statistical significance. Instead, we emphasize the richness of individual insights, highlighting patterns and offering deeper explanatory narratives.


\section{Findings}

We present the findings from our qualitative analysis of social media posts on vibe coding and interviews with vibe coders. Section~\ref{sec:definition} introduces a theory and a definition of vibe coding. Sections~\ref{sec:conversation}--\ref{sec:trust} explore the four core components of our theory: AI interaction, co-creation, flow and trust. Quotes are \emph{\textcolor{mauve}{in dark purple italics}}, and are labeled with their data source (Reddit, Linkedin, or Interview). 
\code{Dark blue italics} denote key code groupings from axial coding (see Section~\ref{subsec:AnalysisMethodology}).

\subsection{Findings---Definition: What is Vibe Coding?}
\label{sec:definition}

When he coined the term in February 2025, Andrej Karpathy defined vibe coding as a ``new kind of coding\ldots where you fully give in to the vibes, embrace exponentials, and forget that the code even exist''.\footnote{As posted on X, \url{https://x.com/karpathy/status/1886192184808149383?lang=en}} 
New terms, however, are not ultimately defined by their originator. Terms evolve \textit{indexically}, meaning their social meaning depends on who uses them and how~\cite{silverstein2003indexical}. This can be especially true of phrases that become memes that are then shaped by who adopts them and how they circulate in online and cultural discourse~\cite{shifman2013memes}. We find that practitioners define vibe coding in a variety of ways. For instance 

\begin{quoteblock}
    \qualquote{Vibe coding is the idea of using LLMs to generate code from natural language}{L33} 

\end{quoteblock}


\begin{quoteblock}
    \qualquote{Vibe coding is\ldots you don't care what the code looks like. All you care is that it behaves the way that you expect it to behave. And if you have an error, you take the error message and you feed it back into the genie.}{I7}
\end{quoteblock}



\begin{figure*}[t]
    \includegraphics[width=0.85\columnwidth]{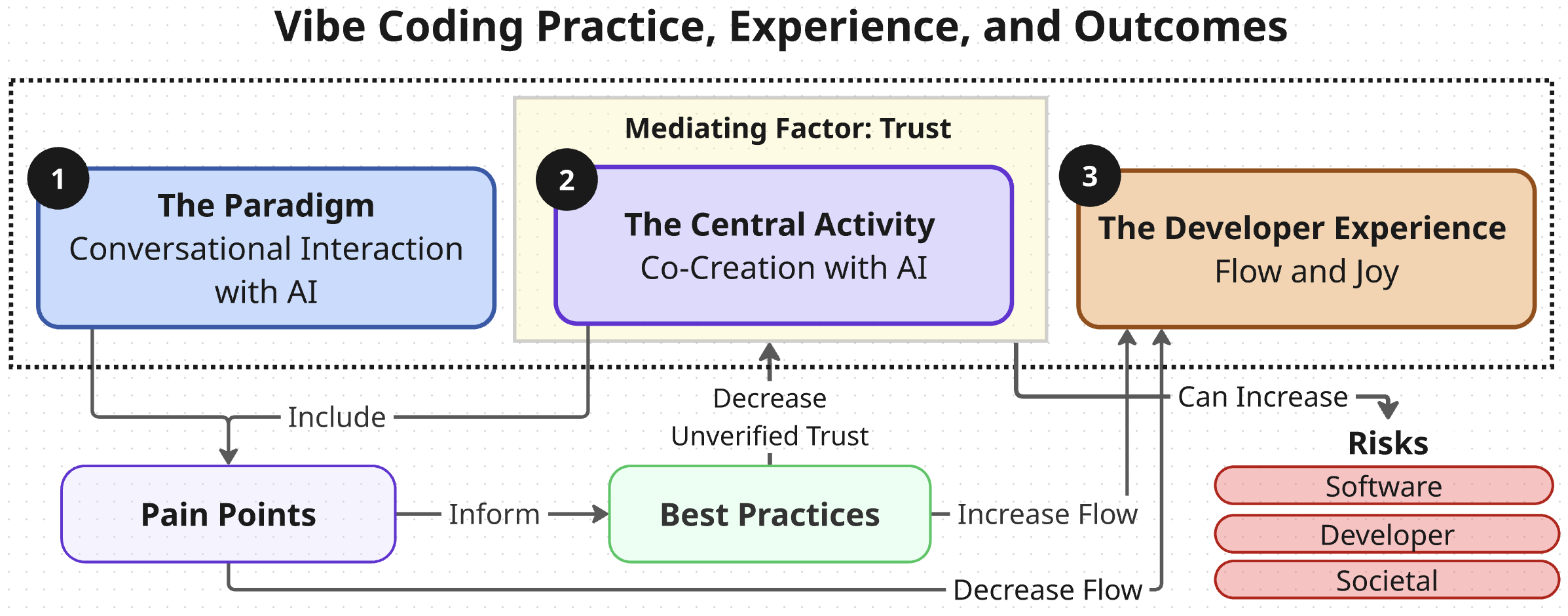}
    \caption{\label{fig:frameworkOverview}Our proposed theory of the vibe coding experience. We identify four core components: \code{Conversational Interaction with AI} (the paradigm), \code{Co-creation with to AI} (the central activity), \code{Flow and Joy} (the developer experience), and \code{AI Trust} (a key enabling and mediating factor). Developers co-create with AI through natural language conversation to achieve flow. \code{Trust in AI} mediates co-creation: increased trust can enable deeper co-creation, thereby enhancing flow. However, increased trust can also lead to increased risks at the software, developer, and societal levels. During both interaction and co-creation, developers experience pain points which can detract from flow. Emerging best practices often ameliorate these pain points, and increase flow. 
    }

    \Description{Image mapping our proposed theory for vibe coding practice and developer experience. We identify four core components: Conversational Interaction with AI (the paradigm), Co-creation with to AI (the central activity), Flow and Joy (the developer experience), and AI Trust (a key enabling and mediating factor). In vibe coding, developers engage with the AI through natural language conversation and delegate tasks to it with the goal of achieving flow. Trust in AI mediates both interaction and delegation: increased trust can enable deeper delegation and smoother interaction, thereby enhancing flow. However, increased trust can also lead to increased risks at the software, developer, and societal level. We find that trust is contextual, shaped by the use case and developer background (e.g., programming experience). Both interaction and delegation have best practices and pain points, which can detract from flow. Emerging best practices are often designed to ameliorate these pain points, and increase flow. However, these best practices can also sometimes programmers to decrease unverified trust (e.g., not delegating as much to AI), thus also decreasing flow.}

\end{figure*}

Despite these varied phrasings, we find that most definitions of vibe coding share key characteristics and constructs. Figure~\ref{fig:frameworkOverview} presents our proposed theory of what practitioners perceive vibe coding to be and how they experience it. We identify four core, interconnected components: \code{Conversational Interactions with AI} (the paradigm), \code{AI Co-creation} (the central activity), \code{Flow and Joy} (the developer experience), and \code{AI Trust} (a key enabling and mediating factor). While we present these four theoretical components as distinct components, it is important to acknowledge that they are closely related and cannot be completely isolated from one another.
Our choice to use these specific component boundaries stems from the need to conduct and present our analysis in a structured and coherent way, despite their interconnectedness.
Grounded in data, we aim to meaningfully convey the complexities of vibe coding practice and experience.


Overall, we find that vibe coding is an emergent, AI-based programming paradigm grounded in natural language interaction and co-creation with an AI agent. It is as much a mindset as a method, prioritizing flow, experimentation, and joy over precision or control. 
Trust in AI acts as a mediating factor for co-creation: greater trust can amplify flow and creative freedom (or, as some practitioners call it, \qualquote{just vibes}{L33}).  As a result, some vibe coders treat the AI as a capable partner and co-creator, even allowing it to make architectural or design decisions with minimal oversight. 

However, this same trust can also increase risk at the software, developer, and societal levels. As a result, some vibe coders prefer to \emph{delegate} tasks to AI rather than co-create. As one interviewee put it, vibe coding exists on \qualquote{a continuum}{I6} with other AI-powered programming paradigms, such as agentic coding. Most commenters agree that vibe coding is a distinct and \qualquote{new way to build}{L81}, they differ in how much trust, interaction, or co-creation required to ``count'' as vibe coding.

These definitional disagreements may stem from the polarized opinions we observed across the three data sources. Some programmers praise vibe coding as (e.g., \qualquote{close to magic}{L18}, or say it \qualquote{brings back the joy of programming}{I7}), while others harshly critique the practice and its practitioners (e.g., \qualquote{So is vibe coding just the dumb-ass version of using AI}{R79}, or \qualquote{Dunning-Kreuger Coders}{R1}). These polarized views call for a deeper exploration of the paradigm, activities, developer experience, and trust dynamics involved.



\subsection{Findings---The Paradigm: Vibe Coding and Conversational Interaction with AI}

\label{sec:conversation}

The first of our four main aspects of vibe coding is: \code{Conversational Interaction with AI}. 
Participants and commenters on Reddit and LinkedIn describe vibe coding as a new programming paradigm, where programming occurs via interaction with AI through \code{conversational natural language}. \qualquote{Vibe coding is when you ask ChatGPT to build software for you---spinning up code, APIs, or entire servers through conversation alone.}{L32} During these natural language conversations with generative AI, the programmer iteratively tries to communicate and refine their intent and goals to the agent. As one interviewee said, \qualquote{vibe coding\ldots is just conveying language through language, what you want to happen, or what you want to be done}{I1}. 

The focus on natural language often extends to how the vibe coder engages with code artifacts. Vibe coding interactions are characterized by \code{little to no code reading and writing}. \qualquote{Ideally, you should not have to interact with the code at all,\ldots you're just guiding the tool}{I8}.
At times when, during traditional programming the programmer would read or modify the program, the vibe coder instead would ask the AI. \qualquote{I think it's basically telling, [the agent]\... what you want to code,\ldots You want to fix something, maybe you'll ask it again instead of, you know, fixing it yourself.}{I10}

Vibe coding also has \code{frequent fine-grained AI interactions}, setting it apart from other AI-assisted programming paradigms. One interviewee contrasted it with agentic programming, explaining, \qualquote{the distinction I make is how much interaction I need to have with the genie\ldots The less I need to interact with the genie, the more agentic it becomes}{I7}. 


\textbf{Paradigm Benefits:} Vibe coders perceive several benefits of high-interaction, natural language programming, including \code{reduced cognitive effort}, \code{reduced unnecessary learning}, \code{increased software development accessibility}, and \code{increased flow and joy}. 
Natural language interactions help \code{reduce cognitive effort} by offloading low-level concerns. \qualquote{Having AI deal with the syntax and specifics frees up brain space for the more important work}{R2}. 
This can also \code{reduce unnecessary learning}, of tools or languages they would rather not learn. One programmer, tasked with making a timer app in React, stated: \qualquote{sometimes I literally do not want anything to do with learning how to program something\ldots  I fucking hate JavaScript in all [its] forms}{R63}. 

As vibe coding often involves limited code reading or writing, it can \code{increase accessibility} to software development for those without formal programming training. For instance \qualquote{vibe coding is dramatically lowering the barrier for non-technical users to create full web apps.}{L6}. Another interviewee described showing this potential to a non-programmer: \qualquote{When I explained\ldots [that] you can just build an app on your phone that's just like every other app, except it does exactly what you tell it to. That was a heavens opening up moment}{I7}. 
This increase in accessibility enables a broader demographic to engage in development, democratizing the ability to innovate and create software.

Finally, natural language programming combined with the short interaction feedback loop can help developers achieve and maintain \code{flow and joy}  
(a defining characteristic of the vibe coding developer experience, see Figure~\ref{fig:frameworkOverview}). We explore this further in Section~\ref{sec:flow}.

\textbf{Interaction Pain Points and Barriers:} Vibe coding AI interaction is not without its challenges. As shown in Figure~\ref{fig:painpoints}, we find that interviewees and commenters report five primary pain points relating to AI interaction and conversation when vibe coding with current platforms and tools. 


First, vibe coders can struggle to \code{accurately specify their intent} to the model as
natural language is inherently imprecise. Mismatched abstractions can lead the model to \qualquote{totally misinterpret[e] what I was saying}{I4}. In addition, minor changes in phrasing or personas can have unexpected results. For example, one commenter noted \qualquote{I put in my rules something along the line of being the best senior programmer\ldots and the Agent stopped doing endpoints and session test and\ldots acted like an arrogant know-it-all dev. That was... a problem}{R43}. Conveying intent can be particularly challenging for those new to software development. 

Vibe coding agents can also suffer from \code{inconsistent conversational memory}, causing the agent to repeat incorrect suggestions and get stuck, resulting in a long, useless \qualquote{prompt spiral}{L51}) where \qualquote{it will just keep re-giving you the same approach, again, and again... Yeah, I would say that's frustrating}{I10}. Vibe coders also report frustration when the agent responds with \code{inaccurate self-assessments} of its ability or past actions. \qualquote{My personal favorite is discovering that what you’re asking relies on essential information from after [its] knowledge cutoff date despite it acting as if it’s an expert on the matter when you ask at the start}{R63}. 

Commenters also note that vibe coding tools such as Claude or Loveable can have \code{slow or costly responses},  
exacerbated by model API rate limits, quotas, and server load. \qualquote{I had issues trying to test Claude with Roo because of these rate limits. It definitely slowed me down}{R75}. As put by one interviewee who regularly vibe codes, \qualquote{just the whole thing needs to be 10 times faster. It doesn't need to be 100 times faster than I can think of, but 10 times faster}{I7}. 

Finally, some vibe coders report overbearing \code{AI company oversight}, where the AI enforces company-imposed behavioral norms or refuses to assist based on tone. One commenter noted \qualquote{if you tell Copilot it isn't listening, it gives you the ``help is available; you're not alone'' suicide spiel. 
}{R63}. Others describe the model refusing requests or scolding users for being impolite: \qualquote{the model said something like ``You cannot speak to me that way. When you have calmed down, let's try again''}{R36}. 
In general, commenters found company-imposed guardrails to be intrusive and frustrating.


All of these pain points can increase frustration and decrease flow. We discuss how these pain points relate to flow, and the best practices developing to overcome them, in Section~\ref{sec:flow}. These pain points also indicate potential directions for future tool support (see our discussion in Section~\ref{sec:discussion}).


\subsection{Findings---The Activity: Vibe Coding and Software Co-Creation}

\label{sec:cocreation}

Interviewees and commenters predominantly characterize vibe coding as true co-creation where rather than solving predefined implementation tasks, the AI often \code{makes higher-order decisions} about features and design. As put by one interviewee, vibe coding involves \qualquote{using an LLM as sort of a co-pilot.\ldots You're actively letting that LLM have a hand in critical high-level decisions\ldots [as] a thought partner}{I11}. 
This co-creation is characterized by \code{AI personification}, with many describing the LLM as a partner or pair programmer, ascribing the LLM agency and autonomy. In vibe coding, \qualquote{the LLM is leading the process, as opposed to you leading the process}{I3}. 


While software co-creation is the primary cited activity in vibe coding, we observe a spectrum from full co-creation to task delegation. 
For some, vibe coding allows the programmer to better focus on design and architecture while delegating implementation-heavy tasks to AI. \qualquote{I've been experimenting with ``vibe coding'', letting AI handle the coding while I focus on steering the ship}{L83}. In addition, for a small number commenters, vibe coding starts the moment the developer uses and trusts AI generated code that they do not understand. As shown in Figure~\ref{fig:frameworkOverview}, we propose that this spectrum between delegation and 
co-creation is primarily mediated by AI trust. We investigate the interaction between co-creation and trust, along with the potential risks of that trust, in Section~\ref{sec:trust}.

\textbf{Co-Creation Benefits:} Vibe coders report experiencing many benefits relating to software co-creation with AI. 
Some vibe coders report that co-creation can \code{enhance learning and brainstorming} when building software, finding benefits in using AI as a sounding board to ask questions. As mentioned by one interviewee, \qualquote{every time I say, I wonder, I have the option of finding out}{I7}. 

Co-creation may also facilitate \code{increased efficiency and productivity}. For one commenter, vibe coding \qualquote{has gotten me to build some pretty complex stuff in 1/4 of the time}{47}. Others found the productivity boost particularly apparent when working with unfamiliar frameworks. During their job as a machine learning engineer, one interviewee talked about vibe coding with a framework that was relatively new to them. \qualquote{Almost all of the PyTest functions I write, I start with having Claude take a crack at it, because\ldots I'm just not fluent enough with that framework yet}{I5}. In these cases, co-creation helps reduce ramp-up time and maintain development momentum.

Finally, many commenters described AI co-creation as a source of substantial joy, and sometimes even addictive. One interviewee noted that vibe coding involves \qualquote{ceding a lot of decision-making authority to the agent}{I6}, making it \qualquote{sort of intoxicating, because normally, you have to make tons of decisions\ldots about everything}{I6}. We explore how co-creation supports \code{flow and joy} in Section~\ref{sec:flow}.

\begin{figure*}[t]
\includegraphics[width=0.90\columnwidth]{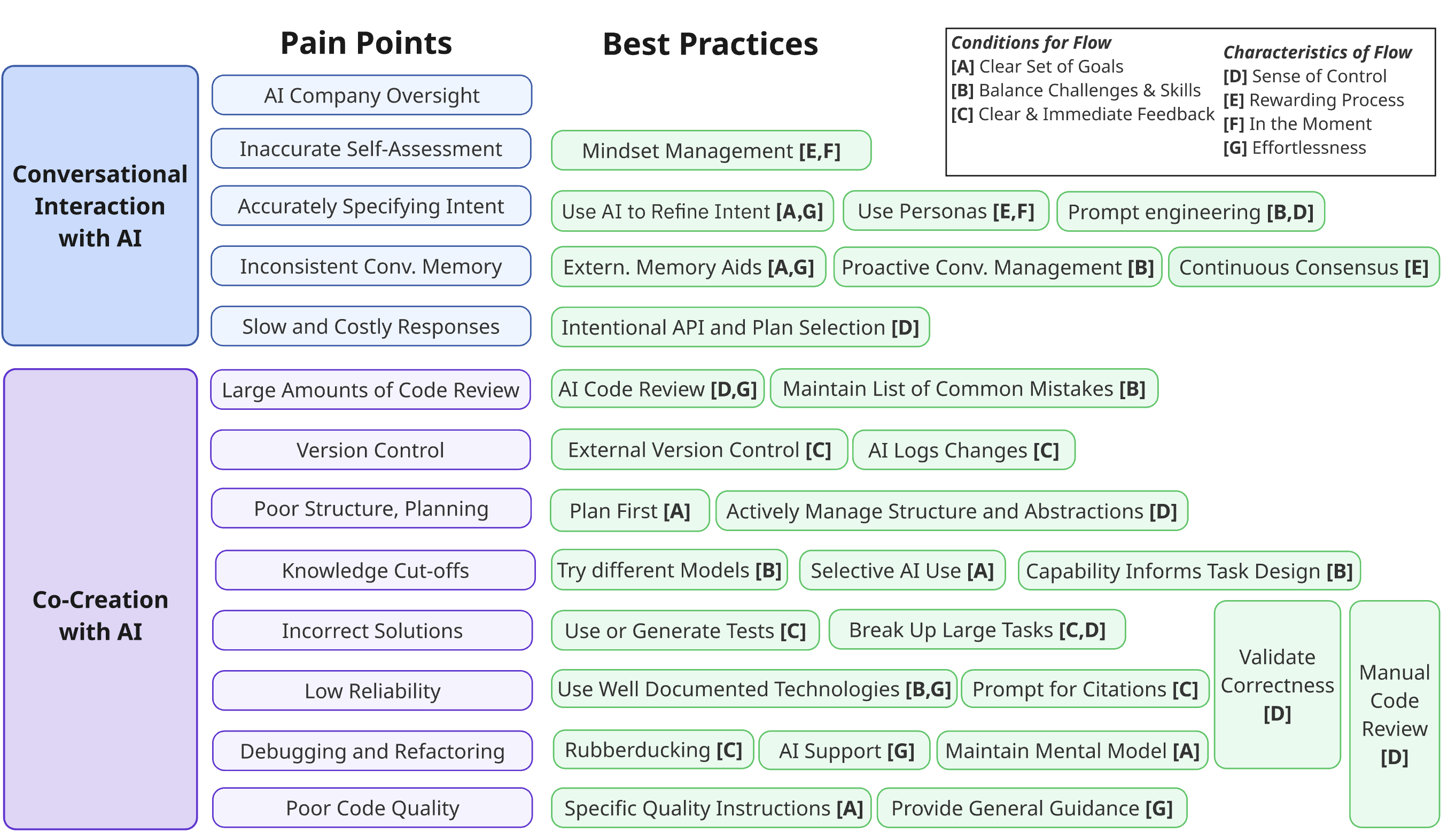}
\caption{\label{fig:painpoints} Mapping between vibe coding pain points, best practices, and characteristics of flow.  Note the best practices not only address the corresponding painpoints (aligned horizontally) but also align with the conditions and characteristics of flow as shown in the top right of the figure.}
\Description{Mapping of vibe coding pain points, best practices, and characteristics of flow. The column on the left lists the pain points. The column on the right lists the best practices. Best practices are additionally mapped to characteristics of flow.}
\vspace{-10pt}
\end{figure*}

\textbf{Co-Creation Pain Points:} However, vibe coding co-creation has its challenges. As shown in Figure~\ref{fig:painpoints}, we identify eight primary co-creation-related pain points for vibe coders.

Co-creation can break down when the vibe coder encounters technical limitations related to \code{knowledge cut-offs}. Some tasks may require working with libraries that were made or modified after the model's training knowledge-cutoff, making co-creation challenging. These issues can be particularly pronounced when working with a private code base or integrating with legacy code systems. \qualquote{since almost every solution we use to common problems is a custom private lib, the LLMs simply have no way of providing value because they know jackshit about my specific issues}{R68}.



Technical limitations can lead to \code{low reliability} during co-creation. 
\qualquote{Planning is boring---until you waste 37+ hours fixing AI hallucinations}{14R}. Vibe coding models make mistakes and provide \code{incorrect solutions}. Sometimes, these responses may also be incomplete or superficial. Some vibe coders report experiencing the model secretly modifying tests or just deleting code without warning. \qualquote{The agent will tell me, like, ``oh, you know, I fixed the tests'' or ``the tests all passed, except for one which isn't our fault''. I'm like, no, it's totally our fault. Like, the tests worked before you started, right?}{I5}. 


A solution may be correct but may lead to \code{poor code quality}, slow or inefficient code, or poor style such as \qualquote{lengthy solution[s] that violate core principles of the language}{R63}. 
Low response quality during vibe coding is particularly apparent with \code{structure and planning}. As reported by one commenter, \qualquote{if the chat is very big, it will forget everything earlier, it will forget any patterns, design and will start to produce bad outputs}{R4}. These structural breakdowns can derail co-creation and increase the burden on the programmer.

Vibe coding also complicates \code{version control}; tools often make large code changes that touch many files, for example: \qualquote{I got too deep in the vibe, took my eye off the ball, and the whole thing spun out of control. I had 30 files in my change log with hours of work uncommitted. It was a \textit{fuckup cascade}}{R35}. These large changes can further lead to challenges with \code{debugging or refactoring};\qualquote{Vibe coding is one thing, vibe debugging is chaos}{R14}. 

Vibe coders try to improve quality through  manual review. However, \code{large amounts of code review} can cause its own problems; \qualquote{I’m more mentally exhausted at the end of the day these days because\ldots I’m working so damn fast and am in constant code review mode}{R43}. 
This can even break the core paradigm, requiring programmers to return to traditional modes of reading and writing code. 
We consider the tradeoffs between trust, flow, and vibe in Section~\ref{sec:trust}.

\subsection{Findings---Developer Experience: Vibe Coding and Flow}

\label{sec:flow}

We find that vibe coding is characterized by a developer experience focused on achieving flow and feeling joy. As put by one commenter, vibe coding is about \qualquote{the art of feeling the flow of a system rather than rigidly adhering to predefined structures}{R41}. As introduced in Section~\ref{subsec:BackgroundFlow}, \textit{flow} is a state of deep, effortless engagement, characterized by intense focus, a sense of control, intrinsic motivation, and merging of action and awareness. Csikszentmihalyi identified three conditions that support flow: clear goals, a balance between challenge and skill, and immediate feedback~\cite{csikszentmihalyi2014flow}. During flow, the process can be just as rewarding as the outcome of the work. 

As shown in Figure~\ref{fig:frameworkOverview}, we propose that interaction and co-creation enable flow and joy in vibe coding. This theme appeared across all data sources. One commenter even used the term \qualquote{flow-code}{41} to describe co-creating while vibe coding. 
Others highlighted how natural language itself fosters flow: \qualquote{I specifically tried to avoid all direct code writing. vibe all the way}{L25}.

This flow state is often accompanied by a feeling of joy. According to an experienced programmer with over 40 years of experience, vibe coding \qualquote{brings back the joy of programming}{I7}, a joy that had been lost through the tedium of software managing dependencies and other software pain points; \qualquote{Oh, you need to update this Library\ldots no, that doesn't work because this other library\ldots So I just gave up. And [that's] part of what makes the Genie magic for me}{I7}. For some, this combination of flow and joy causes vibe coding to not be just a paradigm, but instead \qualquote{a lifestyle}{R13}.

While interaction and co-creation enable flow, current pain points (see Sections~\ref{sec:conversation} and~\ref{sec:cocreation}) can act as barriers to flow, leading to frustration and breaking concentration. 
We find, however, that emerging vibe coding ``best practices''  help re-establish and reinforce the conditions necessary for flow.  Figure~\ref{fig:painpoints} summarizes how specific practices relate to pain points and key flow characteristics. These practices are relevant not only for end users but also for tool designers and researchers seeking to support effective human–AI co-programming. In the rest of this section, we describe these best practices, and the (possibly multiple) pain points they address while enabling flow. 

\textbf{Flow and Interaction Best Practices:} As shown in Figure~\ref{fig:painpoints}, we identified eight best practice categories primarily used by vibe coders to improve conversational AI interactions. We describe these practices, organized by the primary pain point they address, and how they support flow. 

To address challenges in \code{accurately specifying intent}, vibe coders use \code{prompt engineering} tactics and \code{personas} to create clearer, more structured prompts. One interviewee advised treating prompts as instructions to a competent teammate: \qualquote{how much would I need to tell, like, another developer\ldots  who is qualified and competent\ldots but not a world expert?}{I2} These strategies support flow by fostering a sense of control and adjusting the perceived balance between challenges and skills. Some vibe coders also recommend \code{using AI to refine intent}, such as asking about best practices before composing prompts: \qualquote{ask about software security best practices, and then\ldots use the results to create another prompt}{R3}. This can increase flow by increasing both perceived control and effortlessness.

To mitigate \code{inconsistent conversational memory}, some recommend \code{proactive conversation management}, like ending chats when quality drops:  \qualquote{ I 'fire' conversations before they start to lose their mind and start a new one\ldots they give 'hints' that they're losing it}{R1}.  
For those who want a more \textit{effortless} solution to memory mistakes, \code{external memory aids}, such as cursor rules or documentation, can be automatically added to request context. \qualquote{I ended up building a small tool for myself. It generates a code map of the whole project\ldots so AI tools can actually follow what’s going on}{R14}.


To handle \code{slow and costly API responses}, some vibe coders are \code{intentional with API and plan selection}, choosing tools based on task and cost.
\qualquote{Claude 3.7\ldots will use up your Cursor credits whereas Deepseek will not\ldots Depending on the task I will select a specific LLM}{R16}.
Actively making decisions may increase the developer's sense of control over the quality of vibe coding co-creation.

Finally, to manage frustration some coders practice \code{mindset management}. Commenters recommend vibe coders \qualquote{take a Deep Breath}{R35} or use strategies such as persona-based prompting that can make the vibe coding process more rewarding, moment to moment; \qualquote{honestly it’s more for me and keeping my inner monologue in a good head space\ldots while working}{R36}. 

\textbf{Flow and Co-creation Best Practices:} As shown in Figure~\ref{fig:painpoints}, we identified 20 best practices for co-creation. We summarize key strategies aligned with the eight pain points identified in Section~\ref{sec:cocreation}.

When encountering technical limitations such as \code{knowledge cutoffs}, vibe coders will sometimes \code{try different models}, apply \code{selective AI use}, or \code{adapt task design} to better fit model capabilities. \qualquote{I have the most success when designing tasks such that each problem fits inside the context window\ldots It's like having a scalable team of engineers that\ldots need hand holding to tie it all together}{R2}. 
These strategies can lead the developer to experience a better balance between challenge and skill.

To improve code quality, some vibe coders use prompts that include \code{specific quality-focused instructions}, amplifying a sense of control. 
Others prefer a more effortless solution, instead adding documents with  \code{general best-practice guidance} in their context window (e.g., coding style guidelines or internal standards) that can be automatically referenced for all future requests. \qualquote{I gave mine rules for best practices and file formats and other rules for what requirements I need it to follow\ldots it DOES follow all rules}{R75}. 
These two strategies reflect a broader dichotomy in emerging best practices: one favors hands-on, prompt-by-prompt control, while the other seeks effortlessness through one-shot instructions. We hypothesize that developer preference between these two approaches may be related to AI trust. We consider the role of trust in vibe coding in more depth in Section~\ref{sec:trust}. 

To support \code{debugging and refactoring}, vibe coders recommend \code{rubberducking}, building \code{proactive AI-powered workflows}, and maintaining a strong \code{mental model of the codebase}; when debugging, \qualquote{judgement and meta knowledge is key}{R10}. These techniques enhance flow by specifying a clear set of goals, and with rubberducking, fostering a tight, moment-to-moment feedback loop. 

To mitigate for vibe coding's characteristic large and sometimes chaotic code changes, vibe coders recommend using \code{external version control}, as well as \code{asking the AI to log its changes}. \qualquote{Have it write to a file for Git names and version control\ldots Makes it easy to roll back when things go off the rails}{R14}. This explicit tracking offers clear and immediate feedback, supporting flow. 

To combat \code{poor structure and planning}, some commenters recommend  \code{planning first} before vibe coding. This planning phase can involve individual or AI-assisted reflection on potential features, and the desired software architecture. Some vibe coders also recommend \code{actively managing structure and abstractions}, guiding the model toward modular or reusable designs. These practices promote a \textit{clear set of goals} and sustain \textit{a sense of control}, both of which are central to flow-based co-creation.

For \code{low reliability} and \code{incomplete solutions}, best practices include \code{breaking tasks into smaller steps}, and \code{writing or generating tests}. These practices create the conditions necessary for flow by providing \textit{clear and immediate feedback}.  Others recommend choosing \code{well-documented technologies}. \qualquote{AI models are trained on public data. The more common the stack, the better the AI can help you write high-quality code}{R4}. 
Some commenters also emphasize the need to \code{manually validate code} to increase trust in generated code, though others reject this as antithetical to the vibe coding practice/ of avoiding direct code reading (see Section~\ref{sec:conversation}). We discuss the the diverging opinions around code review in \ref{sec:trust}, where we consider trust as a mediating factor of vibe and flow.

\subsection{Findings---Trust as a Mediating Factor of Vibe and Flow}

\label{sec:trust}

In our proposed theory of vibe coding (see Figure~\ref{fig:frameworkOverview}), we identify trust as a key mediating factor that enables co-creation and facilitates flow. Discussions of trust were both explicit (\qualquote{\ldots you can basically set [roo] to auto-approve everything it does if you trust it. It can make prototyping very fast}{R15}), and implicit (\qualquote{Vibe coding is just approving pull requests you don't understand}{R41}).  Trust shapes how much authority a coder is willing to cede to the AI, influencing where they fall on the delegation--co-creation spectrum Section~\ref{sec:cocreation}).
Trust also supports flow by enhancing perceived control and effortlessness~\cite{csikszentmihalyi2014flow}.  However, high trust can also introduce risk. In this section, we explore how trust interacts with vibe coding through a deeper dive on code review during vibe coding, followed by exploring associated risks and how developers regulate trust in context. 

\textbf{Co-creation, Flow, and Trust---A Deep Dive on Vibe Coding and Code Review:} We observe a complex, sometimes conflicting, relationship between best practices, flow, and AI trust when we look more closely at code review of vibe coding. To mitigate for \code{incomplete solutions} and \code{low reliability}, some vibe coders recommend \code{manual review} to increase trust:
\qualquote{The best two tools you have in your toolbox to vibe code well are read the code line by line, and test what the code does}{I11}. However, this strategy is not universal. As noted in Section~\ref{sec:cocreation}, extensive code review is itself a pain point: \qualquote{My 400-line code is now 3000 lines and neither of us can read it anymore}{R63}. Reviewing generated code can be tedious, undermining the very flow and effortlessness vibe coders seek.

As a result, some vibe coders recommend \code{delegating review back to the AI} by asking it to audit its own code. \qualquote{Once you have finished building, take your code and pass it through a leading reasoning model with the following prompt: Please review for production readiness: check for common vulnerabilities,\ldots and ensure adherence to industry best practices}{R41}. This strategy supports flow by preserving a sense of control while enabling effortless review. However, it also signals high trust in model ability; it is unclear how effective these strategies are compared to traditional code review. 

\subsubsection{Vibe Coding Risks} Trust can amplify risks at software, developer, and societal level.
\label{subsubrisks}

\textbf{Risks For Vibe Coded Software:} Commenters report that vibe coding can lead to \code{technical debt}, \code{unmaintainable code}, and \code{buggy or insecure} products. \qualquote{[Vibe coding] can introduce a lot of technical debt, especially if you're not super familiar with, like, the framework or language that it's writing in}{I2}. Security risks are especially concerning: \qualquote{I once tried using ChatGPT to get a simple Spring Boot app\ldots [passwords] got stored in plain text}{R75}.
Some fear that software-related risks make it \code{hard to transition from prototype to product} when relying on vibe coding. \qualquote{Don't try to deploy it\ldots That requires engineering, not vibes. Sorry for gatekeeping but it's true.}{L31}

Reliance on vibe coding may also lead to \code{issues with software team collaboration}. 
For example, one commenter mentioned \qualquote{on the AI team [there] is a, like, prompt engineer. He doesn't have a formal coding background\ldots  sometimes responding to their PRs, I feel like I'm just talking to Claude through a person, which is not efficient for anybody}{I5}.

\textbf{Risks to the Developer:} Commenters also discussed risks for the developer. Some commenters worry that vibe coders who do not critically trust generated code may face \code{legal repercussions} due to their applications leaking sensitive data or ignoring data protection laws. \qualquote{Regulators don’t care about your ``vibes''. Laws don’t care about your feelings}{L3}. Others worry that over-reliance on vibe coding could cause programmers, especially junior programmers, to \code{insufficiently learn key programming concepts}, a risk that \code{can facilitate knowledge-related mistakes} in vibe coded projects. 

Some commenters even worry they are \code{addicted} to vibe coding, a risk that can lead to \code{negative mental health.} \qualquote{If I could go back in time, I would stop myself from using ChatGPT 3.5 for coding\ldots I am literally addicted to it\ldots Whenever I get stuck on a piece of code, I immediately go to AI. It's also a horrible feeling like you are living a lie when people think you're an amazing coder, but it is all AI}{R43}. While we do not perform a thorough investigation of developer-centered risks, we believe that these potential negative side effects of vibe coding is an important direction for future study. 

\textbf{Societal Risks:} Commenters also speculate about potential societal risks. These include;
\code{Climate impact}:\qualquote{I am conflicted over the environmental impacts. Certainly, vibe coding itself, where you just waste the agent's time\ldots and tell it to make a random changes until bugs go away\ldots feels even more wasteful, right?}{I5}; the potential \code{rise in data leaks and scams}: \qualquote{The new way to phish is actually to vibe code an entire functional app. You don't even have to hack anything, you can build a real app as a scammer now}{R3}; and threats to \code{trustworthy OSS}: \qualquote{Used to be you find a GitHub repo with 100 stars, thorough documentation, tests, 100 commits, responsive to bug reports, and you plausibly could trust it. Now it's just some shmo with an LLM who has no idea what they're doing}{R3}. 

\textbf{Calibrating Trust}: Finally, while increased trust can lead to risk, it is important to note that in practice, \textit{trust is contextual}. We find evidence that vibe coders already regulate their trust depending on the specific coding project and its level of risk. Across all data sources, interviewees and commenters distinguish between scenarios when they would or would not vibe code. For example, many commenters report using vibe coding for weekend projects or for building personalized productivity tools \qualquote{Right now I've got like five different tools running on localhost doing things\ldots Not one of them is remotely shippable but all of them were coded with prompts and are solving problems}{R2}. On the other hand, some commenters caution against using vibe coding when dealing with safety critical systems or sensitive data; \qualquote{Soon as you get to password security and data. Or people’s time on the line. Then you gotta think about the serious ramifications}{R2}. This pattern suggests self-regulation and indicates that many developers recognize the trust and risk trade offs in vibe coding and mitigate risk by carefully choosing both when and how they use it. 

%



\subsection{Findings---Summary}
\label{sec:resultssummary}

So far we organized our results around the core tenants of our proposed theory depicted in Figure~\ref{fig:frameworkOverview}: the \textit{paradigm} involving conversational interaction with AI, the \textit{activity} of human-AI co-creation, the developer \textit{experience} centered on flow and joy, and the role of \textit{trust} as a mediating factor. However, while our theory emerged through our analysis, it on its own may not directly answer our initial guiding research questions (see Section~\ref{subsec:rqs}). Here, we summarize our theory and our findings and describe how they directly relate to each initial research question.

\vspace{3pt}

\noindent\textbf{RQ1--\code{Definition}:} \textit{What} is vibe coding? We find that vibe coding is a new paradigm for natural language programming that is involves high level of conversational interaction and co-creation with AI. We further find that the vibe coding developer experience is characterized by flow and joy. Figure~\ref{fig:frameworkOverview} gives an overview of our proposed theory relating AI interaction, co-creation, flow, and trust in a vibe coding context. \vspace{3pt}


\noindent\textbf{RQ2--\code{Practice}:} \textit{Why} do programmers vibe code and \textit{when} do they do it?
We find that vibe coding is often motivated by a desire to achieve flow and joy, decrease cognitive effort, increase productivity, and work fluently with unfamiliar programming tools. Commenters report using vibe coding while building weekend projects, to make custom programming tools for personal use, or for rapid prototyping. Commenters sometimes caution against vibe coding in professional settings, for safety critical systems, or for esoteric or specialized tasks.
\vspace{3pt}

\noindent\textbf{RQ3--\code{Perceptions}:} What are the \textit{perceptions} towards vibe coding?
Perceptions are varied and highly context dependent, depending on the individual's past experiences and trust in AI.

\vspace{3pt}

\noindent\textbf{RQ4--\code{Pain Points}:} What are the \textit{challenges} and \textit{risks} associated with vibe coding? As shown in Figure~\ref{fig:painpoints}, we identify 13 vibe coding pain points experienced by vibe coders. These pain points include challenges with accurately specifying intent, incomplete solutions from models, and dissatisfaction with the large amount of code review. We also identify several risks that may result from vibe coding. These risks can affect vibe coded software, vibe coding developers, or even society at large. 

\vspace{3pt}

\noindent\textbf{RQ5--\code{Best Practice}:} What \textit{best practices} are emerging to handle these challenges? Figure~\ref{fig:painpoints} overviews the many vibe coding best practices that are emerging to address the pain points in vibe coding. Notably, many of these best practices help support and set up the conditions that are necessary for flow (as shown in Figure~\ref{fig:painpoints}). 
    

\section{Discussion and Implications}
\label{sec:discussion}

We now discuss our results and consider potential implications and next steps, focusing on understanding how platform-specific differences may impact our results and understanding how identified benefits risks may play out in practice.

\subsection{Platform specific differences}

In our analysis, we generally consider our results from both social media sites and our interviews holistically, triangulating across all three sources to develop our theory. However, during the analysis process, we did observe some platform-specific differences regarding how vibe coding was perceived and discussed.

LinkedIn posts tended to express more positive sentiment regarding vibe coding, with many posts emphasizing the potential for increased productivity and celebrating the ability to achieve flow. Reddit posts, on the other hand, tended to express more negative sentiments, emphasizing vibe coding pain points and perceived risks. These differences likely reflect the purpose and affordences of each platform, and align with prior findings regarding how people express themselves on social media. For instance, LinkedIn functions as a highly public and professionally oriented platform where users engage in impression management and self-presentation, often emphasizing professional success, optimism, and expertise~\cite{paliszkiewicz2016impression}. In contrast, Reddit users are pseudonymous, affording greater privacy and reducing the social risks associated with expressing criticism or vulnerability. This anonymity, combined with inconsistent forum moderation~\cite{ma2023users} and the psychological tendency for individuals to focus more strongly on negative experiences and emotions~\cite{rozin2001negativity}, contributes to a prevailing negativity bias on Reddit, including on programming-specific forums~\cite{newman2025disclosure}.

Compared to the two social media sites, interviewees tended to voice more nuanced opinions, though individual participants tended to lean more negative or more positive on their vibe coding opinions. This may reflect the broader diversity of developer attitudes towards AI-assisted development tools and workflows.

Overall, while we did observe data-source-specific trends, we note that these trends were not universal; positive, negative, and neutral sentiments toward vibe coding were observed across all three sources. In addition, these differences demonstrate the strength and importance of triangulating across our datasets. All core components of our theory emerged from all three data sources, with only prevalence and sentiment varying between them.



\subsection{Generalized Implications and Future Work}

Is vibe coding a passing trend or the start of a significant paradigm shift? Our study cannot settle this definitively. At a minimum, however, vibe coding has emerged as a distinct sociotechnical practice with its own community, tools, challenges, and best practices, calling attention to potentially important implications for education, practice, tooling, and research. At the same time, we note that the meaning of the term itself appears to be evolving along with the ever-evolving AI-powered software development ecosystem. Regardless, we believe our theory captures core tenants regarding how developers interact with and perceive AI during development. In the rest of this section, we consider broader implications from our findings, with a focus on areas where there is a need for future work.

\noindent{}\textbf{Education:}  If programming increasingly centers on specifying intent in natural language, then curricula should teach students effective GenAI collaboration alongside core CS fundamentals (algorithmic understanding, testing, theory). 
Recent work has already begun exploring how AI-assisted development may reshape computing education, mentorship, and developer learning processes~\cite{feng2026fromjuniortosenior}.
Educators should also address professional upskilling and guard against skill atrophy.

\noindent{}\textbf{Practice:} Participants used vibe coding effectively for personal work, prototyping, and custom tools, but were cautious for production or safety-critical code.
Review load is substantial and risks to quality notable. Teams need processes for this and cannot outsource verification to the same models that generate code.
Sustained flow can, however, bolster morale and motivation.

\noindent{}\textbf{Tooling:} Next-generation tools should address pain points (version-control and provenance integration, longer-horizon conversational memory, reproducible runs) while preserving conditions for flow (responsiveness, a sense of control, low-friction iteration). 
Design for trust calibration and for reviewability/handoffs must also be considered. \qualquote{I think there's some new kind of IDE that's gonna come out that's not gonna look like VS Code at all. And, 1,000 people need to start these ... one that really has the genie baked into the bones of it is gonna be amazing.}{I7}

\noindent{}\textbf{Research:} Examples of next step studies include: (i) empirical tests of reported pain points and real world impact of best practices; (ii) longitudinal studies of skill and trust development with GenAI; (iii) evaluations in industrial settings with complex dependencies and teams; and (iv) systematic security/privacy assessments of vibe-coded artifacts.

\noindent{}\textbf{Risks:} While the societal risks in section \ref{subsubrisks} are  fairly speculative, they are important for contextualizing future directions and research regarding vibe coding. We consider efforts into understanding and mitigating such risks as a promising direction for future work.

\section{Related Work}
\label{rw}






While vibe coding emerged recently, several concurrent studies have begun investigating this paradigm. Studies contrast vibe coding with other AI-assisted approaches, propose formal framings of intent and collaboration, and report empirical observations in lab settings. Our work contributes the first comprehensive qualitative investigation of practitioner experiences and perceptions. In this section we outline recent work based on their connections to our proposed theory, including defining vibe coding, conversational interaction with AI, co-creation with AI, trust as a mediating factor, and flow \& joy.



\subsection{Defining Vibe Coding}
Closest to our work, Sarkar and Drosos analyzed think-aloud recordings of vibe coding sessions to investigate developers' goals, workflows, prompting techniques, debugging approaches and challenges ~\cite{sarkar2025vibecodingprogrammingconversation}.
Their analysis found that vibe coding follows iterative-prompt-and-evaluate cycle: developers alternate between writing natural language prompts, rapidly scanning or testing generated code, and then making occasional manual edits, building trust through iterative verification. The work highlights that conventional programming expertise is still required, but refocused on managing context and verifying AI outputs, indicating that vibe coding augments, rather than replaces, developer skills. The work provides a theory of vibe coding through material disengagement and ``gestalt theory of vibe'', where programmers disengage from direct code manipulation and rely on holistic perception rather than line-by-line analysis. Our work takes a different theoretical approach by grounding vibe coding in flow theory, yet we similarly find that trust functions as the key mediating factor that enables the extent of this disengagement from direct code interaction. 

Sapkota et al. present a comparative review of vibe coding versus a fully autonomous agentic coding paradigm~\cite{sapkota2025vibecodingvsagentic}. 
They characterize vibe coding as a human-in-the-loop conversational mode of development that supports creative exploration, in contrast to agentic coding's goal-driven automation with minimal human intervention. Their taxonomy and use cases indicate that vibe coding excels at early-stage prototyping and educational scenarios, whereas agentic coding shines in enterprise-level tasks like large-scale refactoring and continuous integration. The work argues that future AI-assisted software engineering should harmonize between the two paradigms, combining the strengths of interactive programming and autonomous agents. 

Meske et al. take an intent-driven perspective, formally defining vibe coding as a collaborative human-AI flow, where natural language dialogue replaces code editing, thereby shifting the mediation of developer intent from deterministic instructions to probabilistic inference~\cite{meske2025vibecodingreconfigurationintent}. 
According to their analysis, this reconfiguration redistributes cognitive labor between the developer and the AI, potentially democratizing software creation and accelerating development. However, this also introduces risks such as opaque ``black-box'' codebases. 

Treude and Storey~\cite{treude2025genai} argue that generative AI represents a lasting paradigm shift for empirical software engineering, one where foundational concepts such as ``developer'' and ``coding'' become fluid as AI transitions from a passive tool to an active collaborator. Our study provides the grounded empirical evidence for many of the effects they anticipate, documenting the concrete trust dynamics, pain points, and best practices through which developers actually navigate this shift.


\subsection{Conversational Interaction with AI}

Several studies examined how developers specify, refine, and convey intent through natural language. Tang et al.~\cite{tang2026programmingbychat} mined roughly 75,000 developer messages from over 11,000 IDE chat sessions to study how conversational programming works in practice. They found that new code requests are more rare in comparison to iterative modification, and that developers offload diagnosis and validation to the AI tool. 

Gopalakrishnan describes that fully agentic systems can create a lost-in-the-middle effect, where longer conversations cause context to be lost, or context rot, where outdated or irrelevant information causes a user to lose focus during the coding session ~\cite{gopalakrishnan2026don}. 

Kasibatla et al.~\cite{kasibatla2025aporia} proposed decision-oriented programming, where design decisions become more explicit, persistent objects shared between the developer and their AI tool. They also addressed specification-of-intent as a pain point, similar to our proposed theory.

Geng et al. explore vibe coding in education by observing students using an AI coding platform~\cite{geng2025exploringstudentaiinteractionsvibe}.
They find that students seldom wrote or inspected code directly, where most interactions involved testing or debugging AI-generated prototypes and experienced students crafted richer context-aware prompts.



\subsection{Co-Creation with AI}

This section identifies related work which investigated developers' choices in the co-creation process with AI. Chou et al.~\cite{chou2025buildingsoftwarerollingdice} conducted a grounded-theory study of 20 vibe-coding videos, including live stream coding sessions and opinion videos, and identify a spectrum from full AI delegation to selective co-creation. 

Feng et al. explore how agency is negotiated between junior and senior developers in AI-mediated environments ~\cite{feng2026fromjuniortosenior}. They find that while seniors tend to rely on knowledge and experiences gained without AI, while novices often either over-rely on AI or avoid it entierly.

Kobiella et al.~\cite{kobiella2026throwaway} investigate how GenAI tools reshape product development across skill levels through surveys and interviews with hackathon participants. They find that cloud development environments let even non-technical users produce technically challenging prototypes quickly, but moving from prototype to production requires deeper technical knowledge.

Li et al. present a case study of using an AI-in-the-loop vibe coding approach to rapidly prototype a user interface for data analytics~\cite{li2025usercentereddesignailoop}. Developers and domain experts worked together by conversing with an LLM-based tool to generate interface code from natural language prompts. They found that vibe coding significantly accelerated prototyping and enabled richer feedback in real time. However, the authors also note pitfalls of this approach (e.g., challenges in aligning AI-generated solutions with expert expectations), underscoring that human design expertise and oversight remain critical even when ``coding by vibe''.

Zhang et al.~\cite{zhang2026generativedesign} place the role of vibe coding role within UI prototyping, claiming that generative AI blurs the boundary between design and development into a new style of UI prototyping based on co-creation. This directly relates to our central activity of vibe coding, co-creation with AI.


\subsection{Trust as a Mediating Factor}


This section provides related work on trust as a mediating factor in the vibe coding process, where studies explore what happens when trust is misplaced, how errors may compound, and how developers measure trust.

O'Brien et al.~\cite{obrien2025misconceptions} investigated misconceptions that users hold with LLM coding assistants, distinguishing between tool-specific misunderstandings (e.g. assuming the model can access URLs or remember previous sessions) and deeper misconceptions on LLM structure. They find that users routinely issue underspecified prompts and expect verification which the model may not be able to perform. This connects to challenges we identify, where developers struggle to communicate sufficient context to the AI, and the differences in level of trust developers delegate to AI.

Najem et al.~\cite{najem2025bridging} studied misconceptions users hold about LLM coding assistants through analysis of 500 AI-assisted programming sessions.. They find that people routinely assume the model can access URLs, remember past sessions, or verify code usability and these assumptions in mental models lead to over-delegation and lack of code verification. 

Huang et al.~\cite{huang2025dontvibe} observed 13 developers and surveyed 99 additional developers to understand how experienced practitioners use AI coding agents. They find that these developers overwhelmingly retain rigid control through careful planning before prompting and reviewing each change made. 

Closer to professional team settings, Li et al.~\cite{li2026vibecodingproductdesign} interviewed 22 product team members across enterprises, startups, and academia, and identify tensions between efficiency driven prototyping and reflective design. They also explore trust and responsibility asymmetries within teams that reinforce the collaboration risks we identify in Section~\ref{subsubrisks}.

The shift in agency in vibe coding also carries long-term systemic risks, where Xu et al.~\cite{xu2025ai} find that while AI-assisted tools increase short-term output for less-experienced developers, they significantly increase the maintenance burden on experts, who must review a higher volume of code while seeing a drop in their own original productivity due to accumulated technical debt.

\subsection{Flow and Joy}

We identify flow and joy as one of the defining developer experiences in vibe coding. This section of related work provides concurrent work on the dynamics of flow, what disrupts flow, and how vibe coding reshapes learning.

Jang et al. ~\cite{jang2026evolvingenactions} investigate the professional implications of coding with AI assistants through simulated live coding interviews, where AI assistants change established workflows (such as manual syntax composition) and force developers to find new ways to demonstrate expertise through higher-level planning and rigorous verification of AI-generated output. They explain that rising expectations for speed do not necessarily correlate with expertise, which creates friction in developer experience.

Kusper and Szabo explore vibe coding in educational materials, framing AI as a collaborative "teammate" within project-based learning ~\cite{kusper2025vibecodingeducation}. They suggest that while vibe coding significantly enhances perceived productivity, its impact on student confidence is more closely tied to the experiences of flow and enjoyment. Extending to collaborative settings, Gama et al. observed novice programmers during a vibe coding hackathon ~\cite{gama2026canfeelvibes}. They found that while the paradigm enables rapid prototyping and boosts confidence for newcomers, it often encourages teams to settle on initial AI-influenced ideas without exploring alternative designs or engaging in traditional software engineering practices such as planning, testing, or documentation ~\cite{gama2026canfeelvibes}. 

Royal~\cite{royal2026vibeflow} ran a study with 30 journalism students using AI to build iPhone apps in a mobile development course and mapped their experiences onto Csikszentmihalyi's~\cite{csikszentmihalyi1990flow} eight flow traits\footnote{Citation taken from Royal~\cite{royal2026vibeflow}; we have not read the original source beyond verifying its existence.}. Students reported significant gains in coding confidence, near-universal enjoyment, and vivid descriptions of losing track of time while iterating with AI. Their work demonstrates that the flow dynamics and specification challenges we identify among practitioners also show up with non-traditional coders in educational settings, though their study does not address the code quality, security, and collaboration concerns that emerge in higher-stakes contexts.

Maintaining flow with AI-assisted software engineering remains a technical challenge for which Liu et al. introduced EditFlow ~\cite{liu2026editflow}. Liu et al. noted that despite high benchmark accuracy, AI recommendations often disrupt a developer's natural reasoning process, causing developers to be 19\% longer on tasks due to a disruption of mental flow. 

Across these studies, the literature either 1) contrasts vibe coding with autonomous agents, 2) theorizes its intent-driven mediation, 3) documents local workflows in specific contexts or 4) presents qualitative empirical accounts of practitioners experiences. Our work complements and extends these studies by providing the first systematic analysis of subjective practitioner experiences across diverse contexts, identifying the central role of flow and trust, and synthesizing emerging best-practices for addressing common pain points.

\section{Limitations and Threats to Validity}

Our findings reflect a snapshot window in time (data collected May - July, 2025) and patterns may shift as vibe coding practices evolve.  For social media, our sampling strategy (e.g., emphasizing top/most-recent posts and using a cap) may over-represent high-engagement and early-adopter members of their respective communities. Because platform norms and audience incentives differ (e.g., Reddit's relative anonymity vs. LinkedIn's real-name reputation) we synthesize across data sources and interpret patterns in light of those norms. Our interview pool consisted of 11 individuals, with programming experience in a variety of backgrounds such as web development, data science, machine learning, and back-end development. Their vibe coding experience ranges from personal projects to company projects in their respective areas specific to software engineering. Future work could focus on more targeted populations, such as novice developers or senior developers, to better understand the role of expertise.

Recall bias may have influenced interviewees' responses, as they were asked to describe previous experiences. 
Future work could include in-depth longitudinal or observational studies of developers vibe coding.  Since we collected data based on explicit use of the ``vibe coding'' label, instances of this practice that are not identified with this label may be under-represented.

We did not offer monetary compensation to our interview participants, which may have limited the individuals who participated. Additionally, we recruited participants through snowball sampling, outreach via community Slack channels, mailing lists, and Reddit groups.


\section{Conclusion}

AI use in software development has surged, with developers coining the term ``vibe coding'' to describe AI interaction in development. We provide a qualitative analysis of developer interviews and  Reddit and LinkedIn data, proposing a theory on vibe coding definitions, experiences, and perspectives. 
Our findings suggest that vibe coding may be less defined by technical formalities and rather emphasizes flow, joy, and conversational co-creation with AI. Developers use it for prototyping, personal projects, exploratory work. However, perceptions of vibe coding as a practice remain polarized, with both acceptance and skepticism based on prior experience and trust in AI. 

Vibe coding raises risks such as unclear intent, low quality code, and review fatigue. These challenges extend to reliability, developer education, and trust. 
Developers are adopting best practices that include using tests and planning to supplement AI and reserving vibe coding for personal or low stakes projects. 
The approach signals a shift toward fluid collaboration and creative flow, with the potential to support well being and productivity if balanced against over reliance. 
Future work should examine how vibe coding scales and how best practices can be formalized.

\bibliographystyle{ACM-Reference-Format}
\bibliography{references}

\end{document}